\documentclass[]{pasj01}

\begin{document} 
\Received{}
\Accepted{}

\title{
Hard-tail emission in the soft state of low-mass X-ray binaries  
and their relation to the neutron star magnetic field 
}

\author{Kazumi \textsc{Asai}\altaffilmark{*}}
\altaffiltext{}{MAXI team, RIKEN, 2-1 Hirosawa, Wako, Saitama 351-0198, Japan}
\email{kazumi@crab.riken.jp}
\author{Tatehiro \textsc{Mihara}}
\author{Masaru \textsc{Mastuoka}}
\author{Mutsumi \textsc{Sugizaki}}


\maketitle
\begin{abstract}
Average hard-tail X-ray emission in the soft state of nine
bright Atoll low-mass X-ray binaries containing a neutron star
(NS-LMXBs)
are investigated by using 
the light curves of
MAXI/GSC and Swift/BAT.
Two sources (4U~1820$-$30 and 4U~1735$-$44) exhibit large hardness ratio
(15--50~keV$/$2--10~keV: {\it HR}~$>$~0.1), while
the other sources distribute
at {\it HR}~$\ltsim$~0.1.
In either case, {\it HR} does not depend on the 2--10~keV luminosity.
Therefore the difference of {\it HR} is due to the 15--50~keV luminosity,
which is Comptonized emission.
The Compton cloud is assumed to be around the neutron star.
The size of the Compton cloud would affect the value of {\it HR}.
Although the magnetic field of NS-LMXB is weak,
we could expect a larger Alfv\'{e}n radius than the 
innermost stable circular orbit or the neutron star radius
in some sources. 
In such cases, the accretion inflow is stopped at the Alfv\'{e}n radius
and would create relatively large Compton cloud.
It would result in the observed larger Comptonized emission.
By attributing the difference of the size of Compton cloud
to the Alfv\'{e}n radius,
we can estimate the magnetic fields of neutron star.
The obtained lower/upper limits are consistent with the previous results.
\end{abstract}

\section{Introduction}

A neutron-star low-mass X-ray binary (NS-LMXB)
consists of a weakly magnetized neutron star and
a low-mass star.
NS-LMXBs are known to exhibit soft and hard states in X-ray band,
when their mass-accretion rate is high and low, respectively
(e.g., \cite{Mitsuda1989}; \cite{Matsuoka2013}).
In the soft state,
the spectra are dominated by a soft/thermal component,
while a hard/Comptonized component is also recognized.
The soft/thermal component is well described by
an optically thick standard accretion disk model
\citep{Shakura1973}.
If the magnetic field of the NS is weak ($\le 10^8$~G),
the magnetic field does not to affect the accretion flow,
and the accretion disk can extend down to the NS surface.
Most of the gas in the disk accretes
onto the NS equatorial region.  
As the luminosity decreases, 
a soft-to-hard transition 
occurs in the inner disk
from optically thick to thin.
Then the radius of the inner disk becomes larger.
It is called a truncated disk \citep{Done2007}.
The luminosity of soft-to-hard transition is known to be
1\%--4\% of the Eddington luminosity
\citep{Maccarone2003}.
In the hard state, the spectra are dominated by a hard/Comptonized component,
while a soft/thermal component is still required.
The geometry of the accretion flow changes to the optically thin disk 
accreting to the whole surface of the NS.

The hard/Comptonized component has been observed
in both soft and hard states (see \cite{Barret2001} for a review).
However, there are differences between the two states
in the fitting parameters with the Comptonized component.
In the soft state,
the optical depth ($\tau$) and
the electron temperature ($kT_{\rm e}$)
of the Comptonized component 
are $\tau \sim$~5--15 and $kT_{\rm e}$ = a few keV,
while those in the hard state are 
$\tau \sim$~2--3 and $kT_{\rm e}$ = a few tens of keV. 
Location of the Compton cloud has two possibilities.
One is that the Compton cloud takes place in a transition layer (TL)
(TL: \cite{Titarchuk1998})
located between the disk and NS 
(e.g., \cite{Seifina2012}; \cite{Titarchuk2013}).
The other is an accretion disk corona (ADC)
above the disk (e.g., \cite{Church2014}).
Both the location of the Compton cloud and the origin of the seed photons
remain uncertain.
Furthermore, 
an additional hard X-ray component 
has been detected above $\sim$ 30~keV
in the both states (e.g., \cite{Paizis2006}).
Origin of the emission also remains uncertain,
although there are several models, such as  
Comptonization by bulk motion of accreting matter near the NS
(e.g., \cite{Paizis2006}),
Comptonization by non-thermal electrons accelerated in a jet
(e.g., \cite{DiSalvo2006}), and
synchrotron emission of energetic electrons
(e.g., \cite{Markoff2001}).
Titarchuk, Seifina, and Shrader (2014) reported that
the additional hard X-ray component was evidence for the presence
a hot outer part of the TL
(see also \cite{Seifina2015}).

NS-LMXBs have been classified into two groups,
Z sources and Atoll sources, based on their behavior on the color--color
diagram and hardness--intensity diagram \citep{Hasinger1989}.
Z sources are bright and sometimes reach as close as
the Eddington luminosity.
On the other hand, Atoll sources are less bright,
and some of them exhibit hard/soft spectral state transitions.
In the color--color diagram,
the distribution can be divided into two main regions,
``banana'' and ``island.''
The spectrum is usually softer in the banana than in the island.
The distributions of the diagrams (of color--color and hardness--intensity)
depend on the energy band, the instrument response, and the absorbing column
density towards the source (\cite{Done2003}; \cite{Kuulkers1994}).
Gladstone, Done, and Gierli{\'n}ski (2007) investigated how 
the banana branch moves in the color--color diagram
(soft color: 4--6.4~keV$/$3--4~keV and hard color: 9.7--16~keV$/$6.4--9.7~keV)
for inclination angles of
\timeform{30D},\,  \timeform{60D}, and \timeform{70D}.
The change is mainly in the soft color.
They also investigated the effect of $kT_{\rm e}$
of 2.5, 3.0, and 3.5~keV for each inclination angle. 
This matches very well with the location and the shape of the banana branches
in their color--color diagrams. 

The surface magnetic field of the NS
is believed to be higher in the Z sources ($\sim10^9$~G)
than in the Atoll sources ($\sim10^8$~G)
\citep{Zhang2006}.
For example,
the magnetic field of a Z-source Cyg~X-2 was estimated to
be $2.2 \times 10^9$~G from the observed oscillations in the 
horizontal-branch and the beat frequency model
\citep{Focke1996}.
For a transient Z-source XTE~J1701$-$462,
the magnetic field was estimated to be $\sim$(1--3) $\times 10^9$~G 
from the interaction between the magnetosphere and
the radiation-pressure-dominated accretion disk
\citep{Ding2011}.
Meanwhile, two Atoll sources, 4U~1608$-$52 and Aql~X-1, were estimated to
have magnetic fields of $10^7$--$10^8$~G
from the propeller effect
(4U~1608$-$52: \cite{Chen2006}; \cite{Asai2013},
Aql~X-1: \cite{Campana1998}; \cite{Zhang1998}; \cite{Asai2013}).
However,
Titarchuk, Bradshaw, and Wood (2001)
suggested that both Z sources and Atoll sources have a very low
surface magnetic fields of $\sim10^7$--$10^8$~G based on their
magneto-acoustic wave model and the observed kHz QPOs.
Also using the kHz QPO, 
\citet{Campana2000} estimated the magnetic field of
$\sim$~(1--8)$\times 10^8$~G for Cyg~X-2
from the kHz QPO observability in the variations of the
NS magnetosphere radius (Alfv\'{e}n radius).
\citet{Campana2000} also reported that the magnetic fields
of two Atoll sources,
4U~1820$-$30 and Aql~X-1 were estimated to be  
$\sim 2\times 10^8$~G and $\sim$~(0.3--1)$\times 10^8$~G,
respectively.

In general, the effect of the magnetic fields of NS-LMXB is ignored
because it is weak.
However, in some cases, importance of the effect is suggested
(e.g., \cite{Campana2000}; \cite{Cui1998}; \cite{Ding2011};
\cite{Wang2011}).
\citet{Ding2011} suggested that the inner disk radius could be set by
the magnetospheric radius when the gas pressure from disk
decreases by large radiation pressure and the magnetosphere expands.
\citet{Wang2011} considered a TL between the innermost Keplerian
orbit and the magnetosphere, and suggested that the accretion flow is
disturbed by the several instabilities.
Furthermore, \citet{Campana2000} suggested that 
the disappearance of the kHz QPO was related to the disappearance of
magnetosphere (see also \cite{Cui1998}).
They also noted that the effect of magnetic field of NS must
be taken into account.
In this study, we considered the effect of the magnetic fields
of NS.

We investigated average 
hardness ratio ({\it HR}: 15--50~keV$/$2--10~keV)
in the soft state of nine bright Atoll NS-LMXBs
using MAXI \citep{Matsuoka2009}/GSC
(Gas Slit Camera: \cite{Mihara2011}; \cite{Sugizaki2011})
\footnote{$<$http://maxi.riken.jp/$>$.}
and Swift \citep{Gehrels2004}/BAT
(Burst Alert Telescope: \cite{Barthelmy2005})
\footnote{$<$http://heasarc.gsfc.nasa.gov/docs/swift/results/transients/$>$.}
from 2009 August 15 (MJD = 55058) to 2015 August 15 (MJD = 57249).
In section~2, we describe the data selection and analysis of
the hardness--luminosity diagram.
In section~3, we discuss the relation between hard tail emission and
magnetic fields of NS-LMXB.
The conclusion is presented in section~4. 

\section{Analysis and results}

\subsection{Distribution of hardness ratio (15--50~keV$/$2--10~keV)}

We obtained long-term one-day bin light curves of MAXI/GSC and Swift/BAT
for nine bright Atoll NS-LMXBs.
The light curves of 2--10~keV band of GSC and 15--50~keV band of BAT and
the {\it HR} of the two bands are shown for each source in
figure~1--3. 
The obtained count rates of GSC and BAT were converted to luminosities
by assuming a Crab-like spectrum \citep{Kirsch2005}
and the distances listed in table~\ref{tab1}.
The assumption of Crab-like spectrum is acceptable in the hard state,
because the energy spectrum is dominated by the Comptonized emission
approximated by a power law with the photon index of 1--2.
On the other hand, in the soft state,
the energy spectrum is dominated by the thermal emission.
The obtained luminosity by assuming Crab-like spectrum
is underestimated in the 2--10~keV band, but
is overestimated in the 15--50~keV band.
As a result, the {\it HR} of two energy bands is overestimated
by 2.0 times (see \cite{Asai2015} for detail).
Therefore, we handle only the relative difference
(see appendix).

Figure~\ref{fig4} shows the histograms (number of days)
of the {\it HR} (BAT$/$GSC) of all the nine NS-LMXBs.
The peak around {\it HR}=1 corresponds to the hard state,
while the distribution around {\it HR}=0.1 corresponds to the soft state.
The soft state shows two peaks.
The boundary of the two peaks ({\it HR} at the smallest value)
is 0.09 (dotted line in figure~\ref{fig4}).
We also make the distributions of {\it HR}s for the individual sources
in figure~\ref{fig5}.
The distributions of the soft state are divided into two groups
using the peak value of {\it HR}.
One is that {\it HR} of the peak
is smaller than 0.09: Aql~X-1, 4U~1608$-$52, GX~3$+$1,
GX~9$+$9, GX~13$+$1, and GX~9$+$1.  
The other is that the {\it HR} is larger than 0.09: 4U~1820$-$30 and 4U~1735$-$44.
We note that 4U~1705$-$44 is exceptional since the peak is at {\it HR}=0.09.

\subsection{Hardness--luminosity diagram}

To investigate the difference between two groups in the soft state,
we selected the data of only the soft state.
Table~\ref{tab1} shows periods of the soft state
that are analyzed in this paper.
Four sources (GX~3$+$1, GX~9$+$9, GX~13$+$1, and GX~9$+$1) stayed 
in the soft state during the observed period,
which is obvious since
there is no distribution of the hard state (around {\it HR}=1)
in figure~\ref{fig5}.
In contrast to the four sources,
other five sources show both the soft and hard states.
The individual selection of the soft state is summarized as follows.

\begin{figure*}
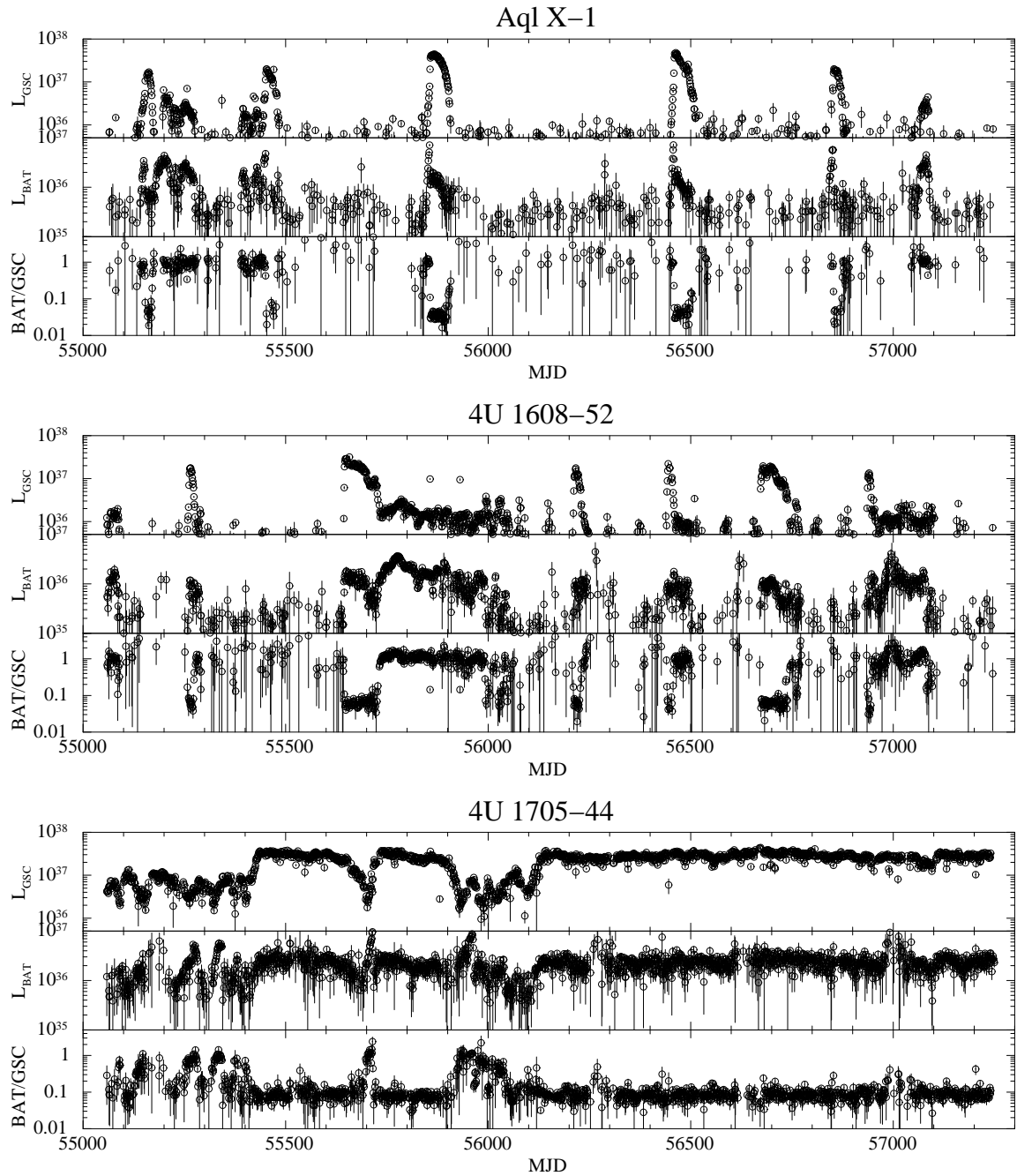

 \begin{center}
 \includegraphics[width=16cm]{fig1-1.eps} 
 \includegraphics[width=16cm]{fig1-2.eps} 
 \includegraphics[width=16cm]{fig1-3.eps} 
 \end{center}
\caption{
One-day GSC light curves in the 2--10~keV band,
one-day BAT light curves in the 15--50~keV band,
and the hardness ratio (BAT/GSC)
of Aql~X-1, 4U~1608$-$52, and 4U~1705$-$44.
$L_{\rm GSC}$ and $L_{\rm BAT}$ are the luminosities in units of erg s$^{-1}$.
Vertical error bars represent 1-$\sigma$ statistical uncertainty. 
}
\label{fig1}
\end{figure*}

\begin{description}

\item[Aql~X-1:]
The source stayed in the soft state during five outbursts (see figure~\ref{fig1}).
The {\it HR} threshold between the soft and hard states was chosen 0.222
in one-day bin data,
which is taken after the analysis in \citet{Asai2015}.
When the state transition continues for more than one day,
we excluded the data and used the data only in a stable {\it HR}.

\item[4U~1608$-$52:]
The source stayed in soft state during six outbursts and 
mini-outbursts during MJD = 55900--56100 (see figure~\ref{fig1}).
The {\it HR} threshold was chosen 0.224 in one-day bin data after \citet{Asai2015}.
We excluded the data during the transition in the same way as Aql~X-1.
We also removed the data of mini-outbursts,
because the spectral states repeated between the soft and hard states
in short durations (less than several tens of days).

\item[4U~1705$-$44:]
The source stays in soft state during three outbursts and 
mini-outbursts during MJD = 55058--55400 and 55800-56100
(see figure~\ref{fig1}).
The {\it HR} threshold was chosen 0.355 in one-day bin data after \citet{Asai2015}.
We excluded
the data during the transition in the same way as Aql~X-1,
and the data of mini-outbursts as 4U~1608$-$52.

\item[4U~1820$-$30:]
Although 4U~1820$-$30 stays in the soft states in most of the time,
it became the hard state ({\it HR}$>$ 0.2: \cite{Asai2015})
eight times (see figure~\ref{fig3}).
We excluded the data in the eight hard states.

\item[4U~1735$-$44:]
Although 4U~1735$-$44 stays in the soft states in most of the time,
it became the hard state ({\it HR}$\sim$1) only once (see figure~\ref{fig3}).
We excluded the data in the hard state.
\end{description}

Figure~\ref{fig6} shows hardness--luminosity diagrams during
the soft state for each source.
The gray backdrop is the distribution of all the nine sources.
The vertical dotted lines represent the
threshold that we determined in figure~\ref{fig4}.
The nine sources are divided into two groups except for 4U~1705$-$44.
One is distributed in the lower part than the threshold of 0.09
(Aql~X-1, 4U~1608$-$52, GX~3$+$1,
GX~9$+$9, GX~13$+$1, and GX~9$+$1.).
The other is distributed in the higher part than the threshold
(4U~1820$-$30 and 4U~1735$-$44).

After excluding 4U~1705$-$44,
both figure~\ref{fig5} and figure~\ref{fig6} show that
eight NS-LMXBs are divided into two groups by the threshold of {\it HR} = 0.09.
This result is consistent with the results of spectral model fitting
for RXTE data of 4U~1820$-$30 and GX~3$+$1 (see Appendix). 
In the next section, we discuss the difference of {\it HR}. 

\section{Discussion}

We investigated average hard-tail X-ray emission in the soft state of
nine bright Atoll NS-LMXBs, where the magnitude of the hard tail
is defined by {\it HR} (15--50$/$2--10~keV, see figure~4 and 5).
We can classify the {\it HR} into at least two groups; larger hard tail 
(4U~1820$-$30 and 4U~1735$-$44) and smaller hard tail (Aql~X-1,
4U~1608$-$52,GX~3$+$1, GX~9$+$9, GX~13$+$1, and GX~9$+$1)
(see figures~6 and 10).
In this section, we give a reasonable phenomenological interpretation
of the observational results. 
The results would be related to properties of each NS,
such as magnetic fields.

\subsection{Differences of inclination, electron temperature, and seed photon}
The difference of {\it HR} (15--50$/$2--10~keV) 
comes from that of the 15--50~keV luminosity, because
the {\it HR} is almost constant for a wide range of 2--10~keV luminosity.
For example, 
the 2--10~keV luminosities of both sources, 
Aql~X-1 and 4U~1820$-$30,
are in the same range of ($\sim$1.2--30)$\times10^{36}$ erg s$^{-1}$.
On the other hand,
the {\it HR} of Aql~X-1 is smaller than 0.09 and that of 4U~1820$-$30
is larger than 0.09 as shown in figure~\ref{fig6}.
The emission in the 15--50~keV energy band is known to originate from
the Comptonized component.

We show the {\it HR} of the energy range (15--50$/$2--10~keV) 
would not be affected by the following differences as below.

\begin{description}

\item[Inclination]
In general, the difference in the {\it HR} distribution
is interpreted by the difference of the inclination of the system
\citep{Gladstone2007}.
When the inclination is high, {\it HR} tend to be large
because of large $\tau$.
GX~13$+$1 is known to be high inclination system because of
the periodic dipping and the deep Fe absorption features
(60--80$^\circ$; \cite{Trigo2012}; \cite{Dai2014}). 
If the distribution of {\it HR} in figure~\ref{fig6}
reflects the inclination, GX~13$+$1 is expected to distribute
in the higher part than the threshold (0.09).
However, the result is opposite.
The other sources would not be high inclination system
because periodic dipping has not been reported. 
Thus the value of {\it HR} would be independent of 
the inclination, when we used the energy band of 15--50$/$2--10~keV.     

\item[Electron temperature $kT_{\rm e}$ of Compton cloud]
In the soft state, $kT_{\rm e}$ is typically a few keV
(see \cite{Barret2001} for a review).
The difference in the {\it HR} distribution
is interpreted by the difference of
$kT_{\rm e}$ of the Compton cloud \citep{Gladstone2007}.
When $kT_{\rm e}$ is high, {\it HR} tend to be large
because Compton-up scattering occurs up to high energy.
Titarchuk, Seifina, and Frontera (2013) reported that 
$kT_{\rm e}$ changes  from 2.9~keV to 6~keV for upper banana
in both 4U~1820$-$30 and GX~3$+$1.
If the distribution of {\it HR} in figure~\ref{fig6}
reflected $kT_{\rm e}$, 
the distributions of {\it HR} in both 4U~1820$-$30 and GX~3$+$1
are expected to be the same.
However, the distributions of the two sources are not the same
and are divided into the top and bottom of the threshold (0.09).
The change of $kT_{\rm e}$ as described by 
Titarchuk, Seifina, and Frontera (2013)
would appear as the scatter in the {\it HR} for each source.

The {\it HR} of 4U~1820$-$30 and 4U~1735$-$44 is larger than the
threshold.
In the case of 4U~1820$-$30, the composition of the accreting matter
is helium \citep{Rappaport1987}.
$kT_{\rm e}$ would be higher than hydrogen accretion of other sources.
However, the orbital period (thus separation) of 4U~1735$-$44
is large and the companion star would be a normal star.
Thus the accreting matter would be hydrogen.
Helium accretion cannot be a common reason for
the large Compton component.
Thus, the classification of two groups of
{\it HR} is independent of $kT_{\rm e}$.

\item[Origin of the seed photon to Compton cloud]
The~Comptonized component is also characterized by the seed photon.
The seed photons is considered to be soft photons from NS surface
and/or disk photons in soft state (e.g., \cite{Sakurai2012};
\cite{Seifina2012}; \cite{Titarchuk2013}).
The difference would not affect the distribution of {\it HR},
because the {\it HR} is almost constant over the large range of luminosity of
(7--50)$\times10^{36}$ erg~s$^{-1}$.
\end{description}

Therefore, we propose that  
the difference between two groups would be due to
the size\footnote{Correctly speaking,
the ``size'' is 
optical depth $\times$ solid angle from the photon source of the Compton cloud.
If the optical depth is similar, the difference of the solid angle
is reflected.
When the distance between the photon source and the Compton cloud
is similar, the solid angle is the proportional to the size of
the Compton cloud.
We use the word ``size'' in this context. 
}
of the Compton cloud.
The group with a larger {\it HR} than 0.09 (4U~1820$-$30 and 4U~1735$-$44)
would have larger Compton cloud than another group with 
a smaller {\it HR} than 0.09.
We notice that the {\it HR} of 4U~1705$-$44 indicates an intermediate
value between two groups.

\subsection{Location of Compton cloud}
In the soft state, Comptonized emission is considered to be formed
in the TL located between the accretion disk and NS surface
(e.g., \cite{Seifina2012}; \cite{Titarchuk2013})
and/or in the ADC above the accretion disk (e.g., \cite{Church2014}).
Although the location is controversial, 
the size of Compton cloud would change with the amount of accretion flow.
However, 
there is no correlation between the observed luminosity and the {\it HR}
between the two groups. 
Here, let us take into account of the magnetic field of the NS.
Although the magnetic field of NS in NS-LMXB is considered to be
weak, in some cases the effect of the magnetic field
is taken into account 
(e.g., \cite{Campana2000}; \cite{Cui1998}; \cite{Ding2011}; \cite{Wang2011}).
In the soft state, 
most of the gas in the disk accretes onto the NS equatorial region,
because the disk can extend to the NS surface or
the innermost stable circular orbit (ISCO). 
The accretion flow is thermalized on the NS surface,
and the emission from there is Comptonized by the plasma around NS.
However, if the magnetic field of the NS is $\ge 10^8$~G,
the Alfv\'{e}n radius would be larger than
the ISCO/NS surface.
In this case, 
the accretion disk would be terminated at the Alfv\'{e}n radius.
The similar situation is suggested by \citet{Ding2011},
although they considered the radiation-pressure-dominated accretion disk.
Then the accretion flow spreading vertically at the Alfv\'{e}n radius
would create relatively large Compton cloud.
In the TL model, the outer boundary of the TL corresponds to 
the Alfv\'{e}n radius.
The difference between {\it HR}s is reflected on
whether the Alfv\'{e}n radius is larger than ISCO/NS surface.

Alfv\'{e}n radius is the radius at
which the gas pressure is equal to the magnetic pressure of the magnetosphere.
It depends on both the mass of the accretion flow
and the magnetic field of the NS.
Using 
the Alfv\'{e}n radius ($R_{\rm A0}$)
in a spherically symmetric mass flow,
the general Alfv\'{e}n radius ($R_{\rm A}$)  
is expressed as the following equation (\cite{Ghosh1979}; \cite{Matsuoka2013}):
\begin{eqnarray}
R_{\rm A} = \eta R_{\rm A0} = 3.7 \times 10^6  \eta (L/10^{36} {\rm erg~s^{-1}})^{-2/7}(B/10^8 {\rm~G})^{4/7} \nonumber \\ 
\times (M_{\rm NS}/1.4\Mo)^{1/7}(R_{\rm NS}/10^6 {\rm cm})^{10/7}~{\rm cm} , \ \ \ \ \ 
\end{eqnarray}
where $G$, $B$, $M_{\rm NS}$, $R_{\rm NS}$, and \Mo ~ are
the gravitational constant, 
the magnetic field of the NS pole,
the mass of the NS, the radius of the NS, and solar mass, respectively.
The original expression (27) 
of Ghosh and Lamb (1979) is changed by substituting
$\mu=BR_{\rm NS}^3$ and  $dM/dt = LR_{\rm NS}/GM$ for the equation of the 
magnetic dipole moment of the NS and the relation
between the mass accretion rate $dM/dt$ and
the total luminosity $L$, respectively.
The factor $\eta $ is expressed as a value of $\sim$ 0.52--1
depending on the model and the effect of
the mass accretion flow.
Approximately,
$\eta \sim 0.52$ corresponds to a disk-like accretion flow and
$\eta \sim 1$ corresponds to a spherical accretion flow\footnote{
We notice that \citet{Wang1997} estimated
$\eta \sim 1$
even in the case of disk-like accretion flow.
They calculated torque in inclined  magnetic moment axis to
the spin axis assuming that the spin axis was normal to the disk plane.
}.

We calculated $R_{\rm A}$
as a function of $L$
when $B$ is from $0.5\times10^8$~G to $3.5\times10^8$~G,
as shown in figure~\ref{fig7}.
We adopted $\eta$ = 0.52 because the accretion flow is considered
to be a disk-like accretion flow in the soft state.
We also adopted 
$M_{\rm NS} = 1.4\Mo$ and $R_{\rm NS} = 10^6 {\rm cm}$.
We also plotted the radius of ISCO and NS.
ISCO is $3r_{\rm g}$, where $r_{\rm g}$ is the Schwarzschild radius
(=$2GM/c^2$).
Here, ISCO is $1.2\times10^6$~cm for $M_{\rm NS} = 1.4\Mo$.
When $L$ is $10^{37}$~erg s$^{-1}$,
$R_{\rm A}$ is larger than ISCO for $B > 1.5\times10^8$~G.
In this case,
the accretion flow would be stopped and spread at the Alfv\'{e}n radius,
and then the relatively large Compton cloud would be created.
As a result, the {\it HR} would be large (for example {\it HR} $>$ 0.09).
On the other hand, when  $B < 1\times10^8$~G, the $R_{\rm A}$
is smaller than both ISCO and NS surface.
In this case, the accretion flow is not affected by the magnetic field,
and  then  most of the gas in the disk accretes
onto the NS equatorial region.
The {\it HR} would be low (for example {\it HR} $<$ 0.09).

Figure~\ref{fig8} illustrates the schematic drawings
of suggesting geometry of NS-LMXB in a soft state.
Assuming $M_{\rm NS}=1.4\Mo$ and $B = 2\times10^8$~G,
(a) $R_{\rm A} = 1.2\times10^6$~cm,
(b) $R_{\rm A} = 1.3\times10^6$~cm,
and (c) $R_{\rm A} = 1.5\times10^6$~cm
for $L = 20\times10^{36}$~erg s$^{-1}$,
$15\times10^{36}$~erg s$^{-1}$, and
$10\times10^{36}$~erg s$^{-1}$, respectively.
According to increasing the luminosity,
$R_{\rm A}$ becomes smaller and the size of Compton cloud
would be smaller.
However, the solid angle from the photon source is constant.
Thus, {\it HR} is constant even if the luminosity is increasing
as long as $R_{\rm A}$ is larger than the ISCO
(figure~\ref{fig8}b and 8c).

Here, we define the threshold luminosity ($L_{\rm th}$)
where the $R_{\rm A}$ equals the ISCO in equation~(1).
In this figure, $L_{\rm th} = 20\times10^{36}$ ~erg s$^{-1}$.
When the luminosity is higher than $L_{\rm th}$,
$R_{\rm A}$ is smaller than ISCO and
{\it HR} is below 0.09 (figure~\ref{fig8}a).
On the other hand, when the luminosity is lower than $L_{\rm th}$,
$R_{\rm A}$ is larger than ISCO and
{\it HR} is above 0.09 (figure~\ref{fig8}b and \ref{fig8}c).

\subsection{Magnetic field derived from the threshold luminosity}
Next, we try to estimate the magnetic field ($B$) of the NS
using the difference of the {\it HR}.
We defined $L_{\rm th}$ in previous subsection.
Here, we use the ISCO rather than the NS surface,
since the ISCO is $1.2\times10^6$~cm which is larger than the $R_{\rm NS}$,
employing traditional values of 
$R_{\rm NS} = 1\times10^6$cm and
$M_{\rm NS} = 1.4\Mo$.
We note that the ISCO is not always larger than the $R_{\rm NS}$
(e.g., \cite{Wang2015}).
Since uncertainty remains for both the radius and mass, 
we adopt the traditional values.
If the luminosity is higher than $L_{\rm th}$,
$R_{\rm A}$ is smaller than the ISCO.
In this case, 
the accretion disk would not be affected by the magnetic field
and
the large Compton cloud would not be created.
The {\it HR} remains small ($<$ 0.09).
When the observed {\it HR} is always smaller than 0.09 in one source,
the minimum observed luminosity becomes the upper limit of $L_{\rm th}$.
On the other hand, when the observed {\it HR} is always larger than 0.09,
the maximum observed luminosity becomes the lower limit of $L_{\rm th}$.
Thus, we can estimate the upper/lower limits of $B$ from
the upper/lower limits of $L_{\rm th}$. 

First, we determined $L_{\rm th}$ from the hardness--luminosity diagrams
(figure~\ref{fig6}).
The value $L_{\rm th}$ are listed in table~\ref{tab2}.
Here, for 4U~1705$-$44, we adopted the lower-end luminosity 
of the distribution on {\it HR} = 0.09
as $L_{\rm th}$, because the {\it HR} of 4U~1705$-$44 distribute 
around {\it HR} = 0.09.
This means that the luminosity change over $L_{\rm th}$.
Substituting $R_{\rm A}$ = ISCO of $1.2\times10^6$~cm and
$L$ = $L_{\rm th}$ in the quotation~(1), 
we obtained $B$ as in table~\ref{tab2}. 
The values $B$ of Aql~X-1 and 4U~1608$-$52 are consistent with those reported
previously.
Although the $B$ of 4U~1820$-$30 ($> 3.2\times 10^8$~G) is somewhat
larger than
the reported value of $\sim 2 \times 10^8$~G \citep{Campana2000},
our values are consistent in the sense
that the $B$ of 4U~1820$-$30 is larger than those of Aql~X-1 and 4U~1608$-$52.

Next, we plotted the
hardness--``normalized luminosity'' diagram for all the sources
in figure~\ref{fig9}b.
The luminosities of figure~\ref{fig6} were normalized by
the $L_{\rm th}$.
Figure~\ref{fig9}a shows the hardness--luminosity diagram for all the sources
to realize the effect of the normalization. 
After the normalization, in figure~\ref{fig9}b,
the data points are distributed on a single trend from 
upper-left to lower-right.
The data points in the upper-left section 
can be interpreted as the accretion flow is spread vertically at $R_{\rm A}$.
Meanwhile, the data points in the lower-right section can be 
interpreted as the accretion flow is not affected by the magnetic fields.
We also presented the hardness--``normalized luminosity'' diagrams
for individual sources in figure~\ref{fig10}. 

According to the fact that
the frequency of the upper kHz QPO 
is related to the Keplerian frequency at the inner edge of an accretion disk, 
\citet{Campana2000} suggested that
the disappearance of kHz QPO at high luminosities may be related to the
disappearance of the magnetosphere
at the time when $R_{\rm A}$ reaches $R_{\rm NS}$.
Then we can assume that the  disappearance of the kHz QPO is related to
a smaller {\it HR} ($<$0.09) in our sources.
In table~\ref{tab2} we listed in existence of reported kHz QPO
\citep{Klis2006}.
The kHz QPO was observed from large {\it HR} ($\ge$ 0.09) sources
(4U~1705$-$44, 4U~1820$-$30, and 4U~1735$-$44)
while it was not observed from small {\it HR} ($\le$ 0.09) sources
(GX~3$+$1, GX~9$+$9, GX~13$+$1, and GX~9$+$1).
In the cases of Aql~X-1 and 4U~1608$-$52,
the kHz QPO was detected even though they have small {\it HR}.
For Aql~X-1,
\citet{Campana2000} reported that
the kHz QPO disappeared at the luminosity (2--10~keV)
higher than $3.6 \times 10^{36}$~erg s$^{-1}$. 
Similarly,
for 4U~1608$-$52, 
the kHz QPOs were detected 
only  when the luminosity (2--20~keV)
dropped down to $10^{37}$~erg s$^{-1}$ \citep{Mendez1998}.
Namely, the kHz QPO was not detected in the soft state ($> 10^{37}$~erg s$^{-1}$). 
Therefore, the existence/absence of the kHz QPO is consistent with
our results of the large/small {\it HR}.

\section{Conclusion}

We investigated average hard-tail X-ray emission in the soft state of
nine bright Atoll NS-LMXBs by using the light curves of
MAXI/GSC and Swift/BAT, and 
tried to relate the difference of {\it HR}s to that of $B$ of NSs.
When $R_{\rm A}$ is larger than the ISCO, 
the accretion disk would be stopped and spread at $R_{\rm A}$,
and then the relatively large Compton cloud would be created.
Then, the {\it HR} would become large ($>$ 0.09), and
the observed luminosity is lower than the $L_{\rm th}$,
where the $L_{\rm th}$ is the luminosity at which the $R_{\rm A}$ equals the ISCO.
We can drive the lower limit of $B$ of the NS
from the lower limits of $L_{\rm th}$. 
Since {\it HR} of 4U~1820$-$30 and 4U~1735$-$44 is large,
$B$s is estimated as $B$$\gtsim 2.5 \times10^8$~G.
The upper limit of $B$ is given for other seven sources.
These results are consistent with the previous results.

\begin{ack}
We would like to acknowledge the MAXI team for MAXI operation
and for analyzing real-time data.
\end{ack}

\appendix
\section*{Hardness--intensity diagram}

Here we investigate the validity of assuming a Crab-like spectrum
for a soft state in the conversion from the count rate of GSC and BAT
to the luminosity (see subsection~2.1). 
In a soft state, the luminosity obtained in this way
is underestimated in the
2--10~keV, but is overestimated in the 15--50~keV.
Therefore, we handle only the relative difference of {\it HR}.
Here we compare {\it HR} 
obtained by assuming a Crab-like spectrum
with that obtained by spectral model fitting for RXTE data.
The latter was taken from
table~4 of Titarchuk, Seifina, and Frontera (2013) for 4U~1820$-$30 and
table~4 of Seifina and Titarchuk (2012) for GX~3$+$1.
Figure~\ref{fig11} shows the hardness--luminosity diagram
in both cases.
The data of 4U~1820$-$30 included those of the soft and hard state
in both cases.
The energy band of {\it HR} is slightly different.
The {\it HR} of figure~\ref{fig11}a is 15--50~keV$/$2-10~keV.
In figure~\ref{fig11}b, the {\it HR}
of 4U~1820$-$30 and GX~3$+$1 is 10--50~keV$/$3--10~keV and
10--60~keV$/$3--10~keV, respectively.
In spite of the difference of energy band,
the {\it HR} of 4U~1820$-$30 is larger than that of GX~3$+$1
in both figure~\ref{fig11}a and \ref{fig11}b.
Thus, relative difference of {\it HR} assuming a Crab-like spectrum
is consistent with the result of spectral model fitting.

\clearpage

\begin{table*}
  \tbl{Periods of the soft states defined by MAXI/GSC and Swift/BAT,
and distance used in this paper.}{%
  \begin{tabular}{llcc}
      \hline
      Name & Period of soft states (MJD) & Distance (kpc) & Ref.$^*$\\ 
      \hline
      Aql~X-1 & 55157--55170, 55451--55479, 55857--55899,
               56460--56513, 56853--56873 & 5 & (1)\\
      4U1608$-$52 & 55260--55272, 55645--55721, 56213--56226,
                   56442--56445, 56673--56734, 56937--56945 & 4.1 & (2)\\
      4U1705$-$44 & 55410--55684, 55717--55915, 56048--57249 & 7.4 & (1) \\   
      GX~3$+$1 & 55058--57249  & 4.5 & (4)\\
      GX~9$+$9 & 55058--57249  & 5 & (3) \\
      GX~13$+$1 & 55058--57249 & 7 & (1)\\
      GX~9$+$1 & 55058--57249  & 5 & (1)\\
      4U~1820$-$30 & 55058--55288, 55306--55615, 55629--56130, 
                  56141--56474, 56482--56619, 56681--56986 & 7.6 & (1)\\
                 & 56995--57132, 57145--57151, 57167--57249 \\
      4U~1735$-$44 & 55058--56766, 56775--57249 & 8.5 & (2)\\
      \hline
    \end{tabular}}
\label{tab1}
\begin{tabnote}
\footnotemark[$*$]
(1) \citet{Liu2007}, (2) \citet{Galloway2008},
(3) Christian and Swank (1997), and (4) Kuulkers and van der Klis (2000).
\end{tabnote}
\end{table*}

\begin{table*}
  \tbl{Threshold luminosity and corresponding magnetic fields.}{%
  \begin{tabular}{lcccccc}
      \hline
      Name & {\it HR} of peak & $L_{\rm th}$$^*$ & $B$ & kHz QPO$^\dagger$ & Reported $B$ & Ref.$^\ddagger$\\ 
           &    & $10^{36}$ erg s$^{-1}$ & $10^{8}$ G & & $10^{8}$ G & \\
      \hline
      Aql~X-1       & $<$ 0.09 & $\ltsim 11$ & $\ltsim 1.4$ & $\circ$ & 0.3--1.9 & (1), (2), (3), (4)\\
      4U1608$-$52   & $<$ 0.09 & $\ltsim 8$ & $\ltsim 1.2$ & $\circ$ & 0.5--1.6 & (4), (5) \\
      4U1705$-$44   & $=$ 0.09 & $\ltsim 20$ & $\ltsim 1.9$ & $\circ$ & -- & --\\
      GX~3$+$1      & $<$ 0.09 & $\ltsim 8$ & $\ltsim 1.2$ & -- & -- & --\\
      GX~9$+$9      & $<$ 0.09 & $\ltsim 9$ & $\ltsim 1.3$ & -- & -- & --\\
      GX~13$+$1     & $<$ 0.09 & $\ltsim 30$ & $\ltsim 2.4$ & -- & -- & --\\
      GX~9$+$1      & $<$ 0.09 & $\ltsim 23$ & $\ltsim 2.1$ & -- & -- & --\\
      4U1820$-$30   & $>$ 0.09 & $\gtsim 55$ & $\gtsim 3.2$ & $\bullet$ & $\sim 2$ & (3) \\
      4U1735$-$44   & $>$ 0.09 & $\gtsim 34$ & $\gtsim 2.5$ & $\bullet$ & -- & -- \\
      \hline
    \end{tabular}}
\label{tab2}
\begin{tabnote}
\footnotemark[$*$] Threshold luminosity in the 2--10~keV band (see text for explanation).  \\
\footnotemark[$\dagger$]
$\bullet$: kHz QPOs have been reported in soft state. 
$\circ$: kHz QPOs have been reported, although the state is not clear.
--: not reported \citep{Klis2006}.  \\
\footnotemark[$\ddagger$] (1) propeller effect \citep{Campana1998},
(2) propeller effect \citep{Zhang1998},
(3) kHz QPO \citep{Campana2000},
(4) propeller effect \citep{Asai2013}, and
(5) propeller effect \citep{Chen2006}.
\end{tabnote}
\end{table*}

\begin{figure*}
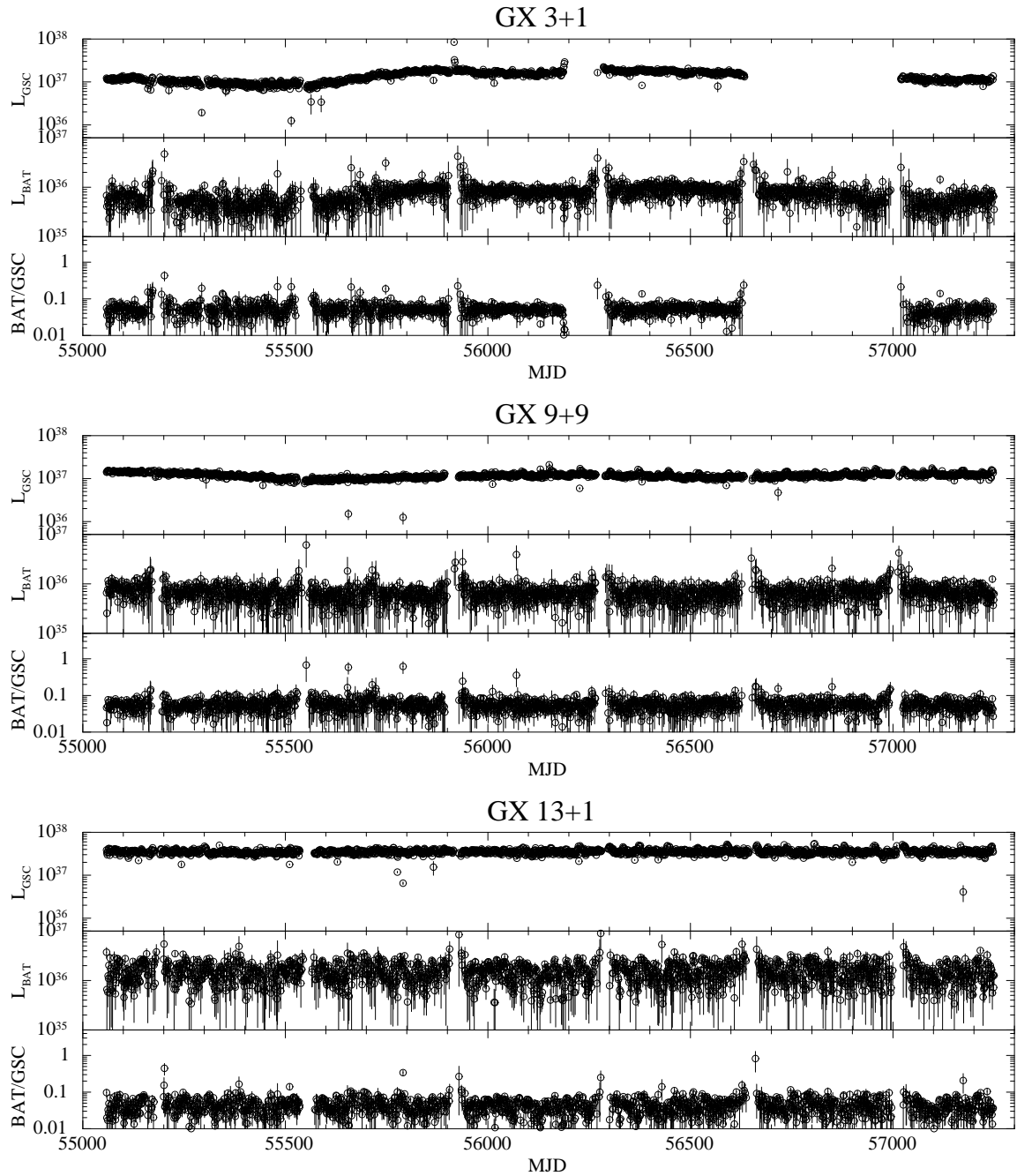

 \begin{center}
 \includegraphics[width=16cm]{fig2-1.eps} 
 \includegraphics[width=16cm]{fig2-2.eps} 
 \includegraphics[width=16cm]{fig2-3.eps} 
 \end{center}
\caption{Same as figure~\ref{fig1}, but for GX~3$+$1, GX~9$+$9, and GX~13$+$1.
One-year periodic rises before and after data gaps in the Swift/BAT light curves
are due to instrumental effect.
MAXI/GSC data gaps in GX~3+1 in MJD=56187--56287 and 56652--56997
are due to contamination source.
}
\label{fig2*}
\end{figure*}

\begin{figure*}
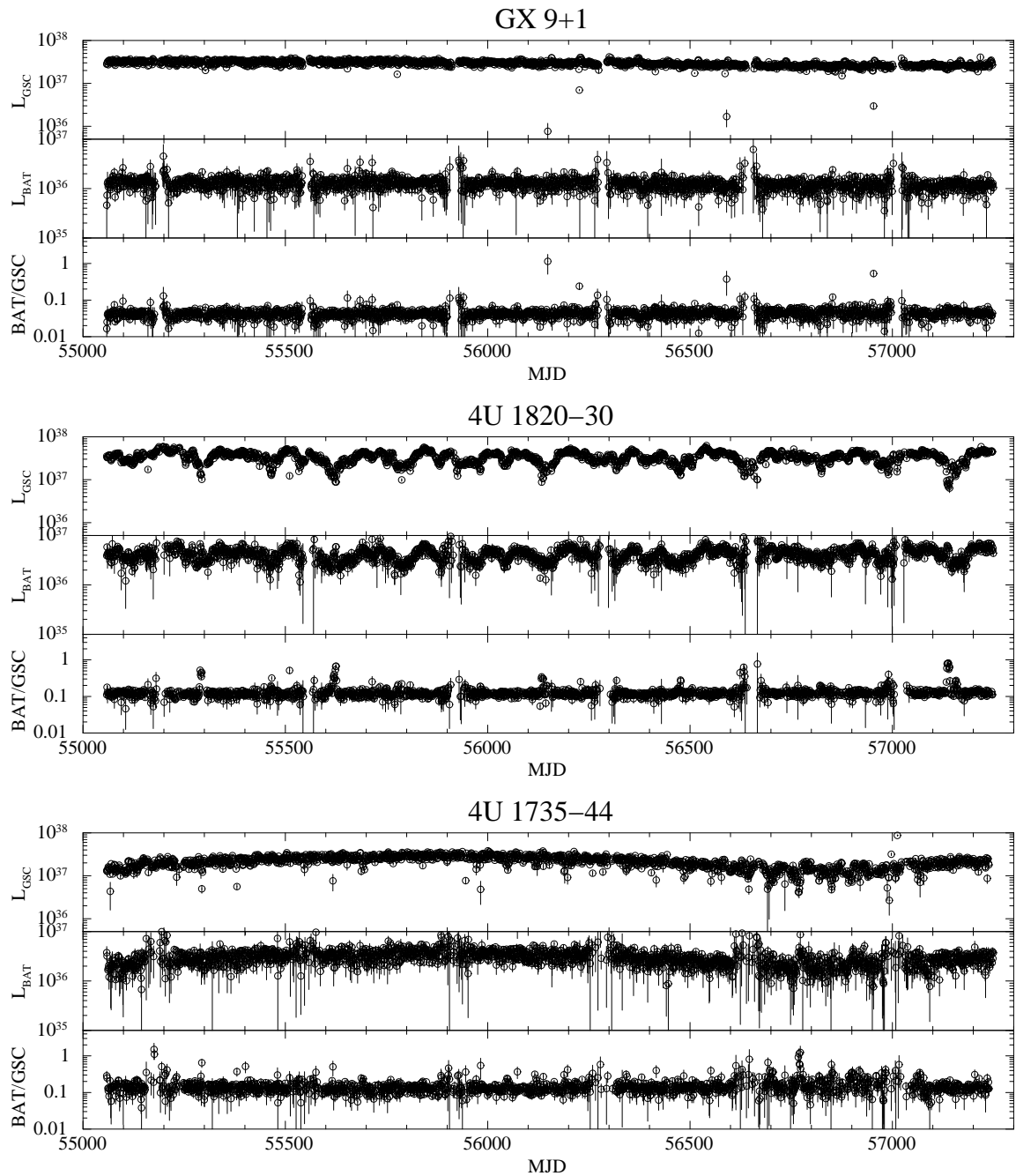

 \begin{center}
 \includegraphics[width=16cm]{fig3-1.eps} 
 \includegraphics[width=16cm]{fig3-2.eps} 
 \includegraphics[width=16cm]{fig3-3.eps} 
 \end{center}
\caption{
Same as figure~\ref{fig1}, but for 
GX~9$+$1, 4U~1820$-$30, and 4U~1735$-$44.
}
\label{fig3}
\end{figure*}

\begin{figure*}
 \begin{center}
  \includegraphics[width=8cm]{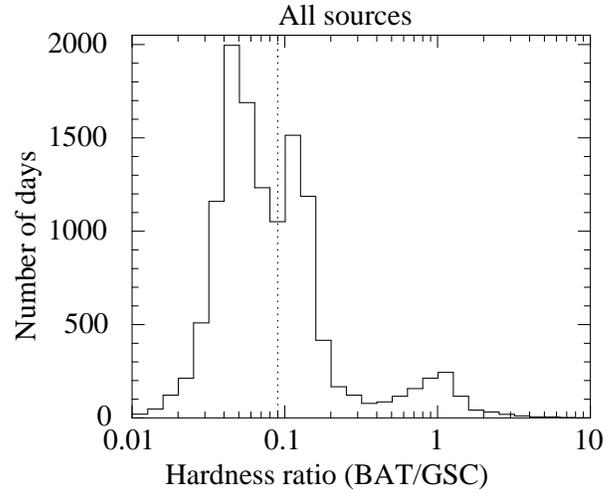}
 \end{center}
\caption{Distributions of hardness ratios of BAT/GSC for all nine NS-LMXBs.
The distributions were constructed from data 
with a significance $>$ 1~$\sigma$.
The vertical dotted line represented the threshold between
two soft states (see text for explanation).
}
\label{fig4}
\end{figure*}

\begin{figure*}
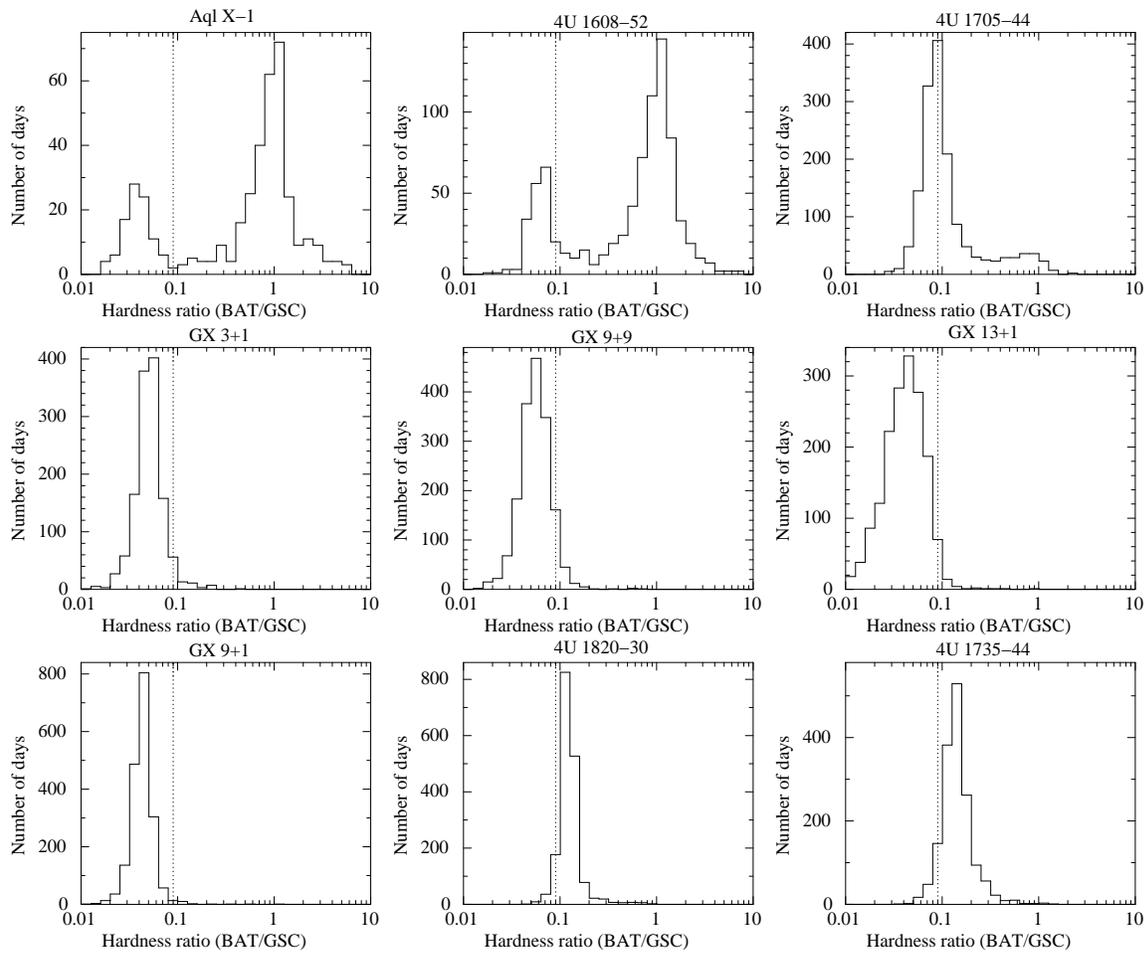

 \begin{center}
  \includegraphics[width=5cm]{fig5-1.eps} 
  \includegraphics[width=5cm]{fig5-2.eps}
  \includegraphics[width=5cm]{fig5-3.eps}
  \includegraphics[width=5cm]{fig5-4.eps}
  \includegraphics[width=5cm]{fig5-5.eps}
  \includegraphics[width=5cm]{fig5-6.eps}
  \includegraphics[width=5cm]{fig5-7.eps}
  \includegraphics[width=5cm]{fig5-8.eps}
  \includegraphics[width=5cm]{fig5-9.eps}
 \end{center}
\caption{Distributions of hardness ratios of BAT/GSC for nine NS-LMXBs.
The vertical dotted lines represented the threshold that we determined in 
figure~\ref{fig4}.}
\label{fig5}
\end{figure*}

\begin{figure*}
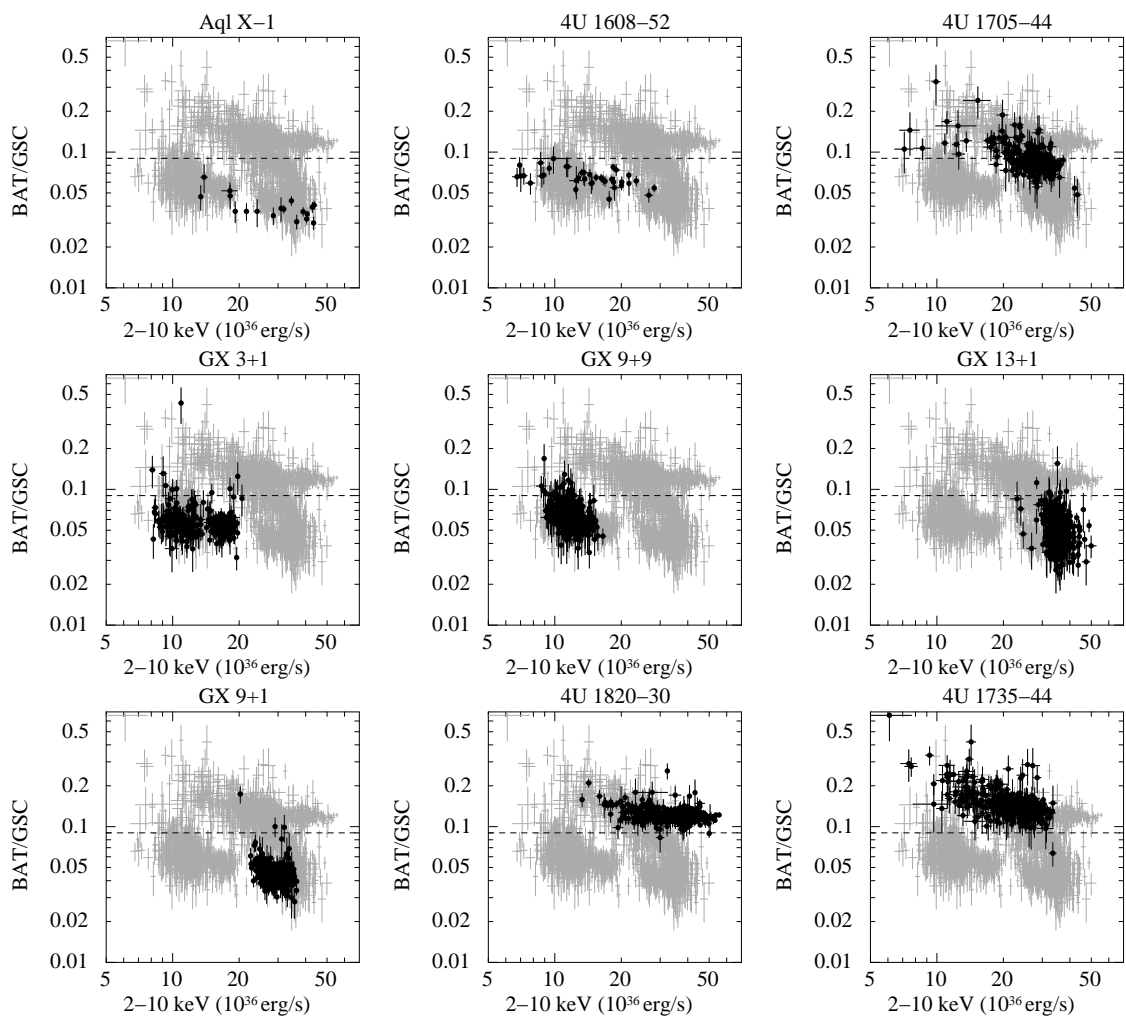

 \begin{center}
  \includegraphics[width=5cm]{fig6-1.eps} 
  \includegraphics[width=5cm]{fig6-2.eps}
  \includegraphics[width=5cm]{fig6-3.eps}
  \includegraphics[width=5cm]{fig6-4.eps}
  \includegraphics[width=5cm]{fig6-5.eps}
  \includegraphics[width=5cm]{fig6-6.eps}
  \includegraphics[width=5cm]{fig6-7.eps}
  \includegraphics[width=5cm]{fig6-8.eps}
  \includegraphics[width=5cm]{fig6-9.eps}
 \end{center}
\caption{Hardness--luminosity diagram,
which plotted against a backdrop of all nine NS-LMXBs.
The data are 5-d averaged.
The distributions are constructed from data 
with a significance $>$ 3~$\sigma$.
The horizontal dotted lines represented the the
threshold that we determined in figure~\ref{fig4}.}
\label{fig6}
\end{figure*}

\begin{figure*}
 \begin{center}
  \includegraphics[width=8cm]{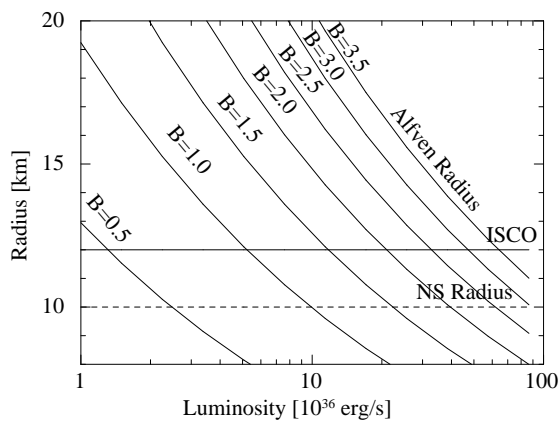}
 \end{center}
\caption{Alfv\'{e}n radius ($R_{\rm A}$)
as a function of luminosity,
where $B$ is from $0.5\times10^8$~G to $3.5\times10^8$~G.}
\label{fig7}
\end{figure*}

\begin{figure*}
 \begin{center}
  \includegraphics[width=8cm]{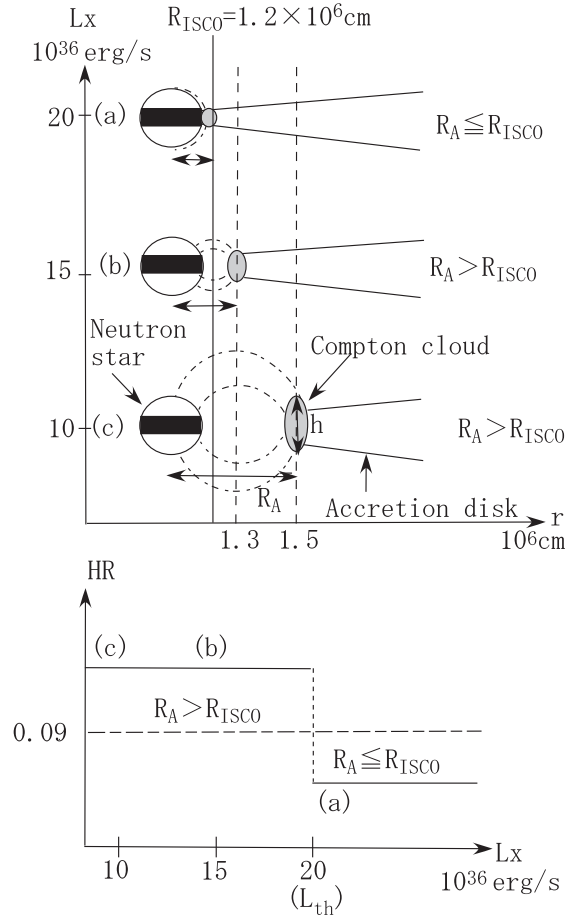}
 \end{center}
\caption{Schematic drawings of suggesting geometry of NS-LMXB in the soft state.
Assuming $M_{\rm NS}=1.4\Mo$ and $B = 2\times10^8$~G, 
(a) $R_{\rm A} = 1.2\times10^6$~cm, (b) $R_{\rm A} = 1.3\times10^6$~cm,
and (c) $R_{\rm A} = 1.5\times10^6$~cm
corresponding to $L = 20\times10^{36}$,
$15\times10^{36}$, and
$10\times10^{36}$~erg s$^{-1}$, respectively.
The {\it HR} representing the solid angle of the Compton cloud
changed between (a) and (b), while it stays constant in (b) and (c).
}
\label{fig8}
\end{figure*}

\begin{figure*}
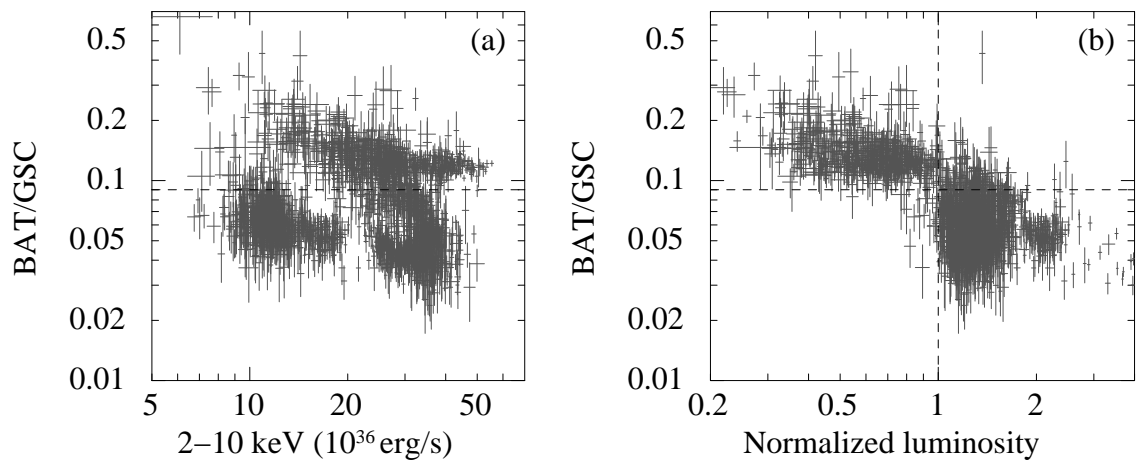

 \begin{center}
  \includegraphics[height=6cm]{fig9-1.eps}
  \includegraphics[height=6cm]{fig9-2.eps}
 \end{center}
\caption{(a) Hardness--luminosity diagram
for all the nine NS-LMXBs, which is the gray backdrop in figure~\ref{fig6}.
(b) Hardness--normalized luminosity diagram
for all the nine NS-LMXBs (see text for explanation).
The horizontal dotted lines in both panels
represent the threshold that we determined
in figure~\ref{fig4}.}
\label{fig9}
\end{figure*}

\begin{figure*}
 \begin{center}
  \includegraphics[width=5cm]{fig10-1.eps} 
  \includegraphics[width=5cm]{fig10-2.eps}
  \includegraphics[width=5cm]{fig10-3.eps}
  \includegraphics[width=5cm]{fig10-4.eps}
  \includegraphics[width=5cm]{fig10-5.eps}
  \includegraphics[width=5cm]{fig10-6.eps}
  \includegraphics[width=5cm]{fig10-7.eps}
  \includegraphics[width=5cm]{fig10-8.eps}
  \includegraphics[width=5cm]{fig10-9.eps}
 \end{center}
\caption{Hardness--normalized luminosity diagram,
which plotted against a backdrop of all the nine NS-LMXBs
(figure~\ref{fig9}b).
The data are 5-d averaged.
The distributions are constructed from data 
with a significance $>$ 3~$\sigma$.
The horizontal dotted lines represent the threshold that we determined 
in figure~\ref{fig4}.}
\label{fig10}
\end{figure*}

\begin{figure*}
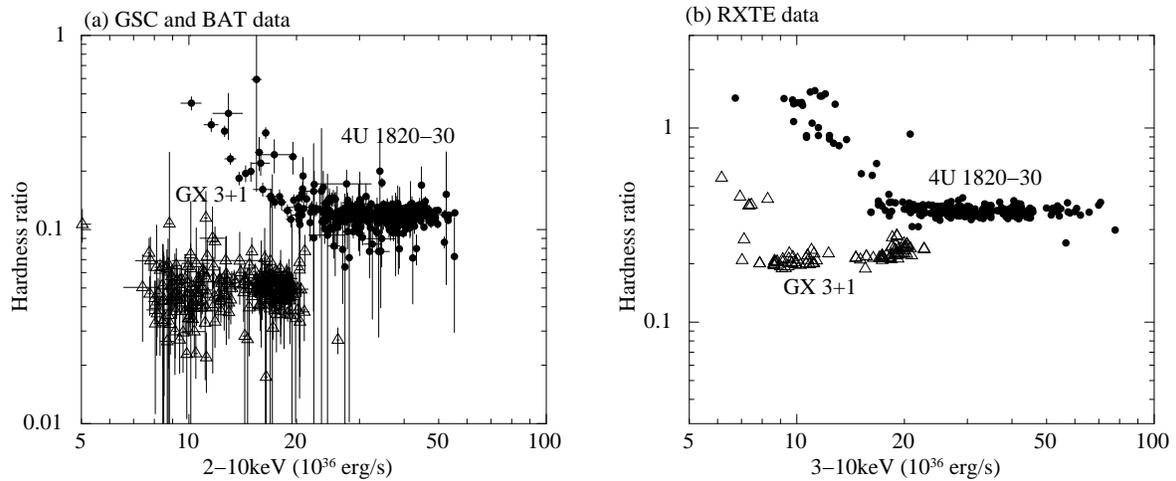

 \begin{center}
  \includegraphics[width=8cm]{fig11a.eps}
  \includegraphics[width=8cm]{fig11b.eps}
 \end{center}
\caption{Hardness--luminosity diagram of 4U~1820$-$30 and GX~3$+$1.
The data of 4U~1820$-$30 included the data of both soft and hard state.
(a) The data are obtained by GSC and BAT.
The 2--10~keV luminosities are obtained by assuming Crab-like spectrum.
The {\it HR} is 15--50~keV$/$2-10~keV. 
The data are 5-d average.
(b) The data are obtained by RXTE, which is listed in table~4
of Titarchuk, Seifina, and Frontera (2013) for 4U~1820$-$30 and
table~4 of Seifina and Titarchuk (2012) for GX~3$+$1.
The {\it HR} of 4U~1820$-$30 and GX~3$+$1 is
10--50~keV$/$3--10~keV and 10--60~keV$/$3--10~keV, respectively.}
\label{fig11}
\end{figure*}

\end{document}